\documentclass[aps,prb,amsmath,amssymb,amsfonts,notitlepage]{revtex4-1}
\usepackage{graphicx}
\usepackage{color}

\begin{document}

\title{Interface control of electronic transport across the magnetic phase transition in SrRuO$_3$/SrTiO$_3$ heterointerface}
\author{S. Roy$^1$, C. Autieri$^2$, B. Sanyal$^2$, T. Banerjee$^1$}

\email{T.Banerjee@rug.nl}

\affiliation{$^1$ Physics of Nanodevices, Zernike Institute for Advanced Materials, University of Groningen, Groningen, The Netherlands \\
$^2$ Department of Physics and Astronomy, Uppsala University, Box-516, 75120 Uppsala, Sweden}

\begin{abstract}

{\bf The emerging material class of complex-oxides, where manipulation of physical properties lead to new functionalities at their heterointerfaces, is expected to open new frontiers in Spintronics. For example, SrRuO$_3$ is a promising material where external stimuli like strain, temperature and structural distortions control the stability of electronic and magnetic states, across its magnetic phase transition, useful for Spintronics.
Despite this, not much has been studied to understand such correlations in SrRuO$_3$. Here we explore the influence of electron-lattice correlation to electron-transport, at interfaces between SrRuO$_3$ and Nb:SrTiO$_3$ across its ferromagnetic transition, using a nanoscale transport probe and first-principles calculations. We find that the geometrical reconstructions at the interface and hence modifications in electronic structures dominate the transmission across its ferromagnetic transition, eventually flipping the charge-transport length-scale in SrRuO$_3$. This approach can be easily extended to other devices where competing ground states can lead to different functional properties across their heterointerfaces.}

\end{abstract}

\maketitle

\section*{Introduction}

 Complex oxides have propelled a vast research field, in establishing itself as next frontiers in electronic materials by offering unique prospects to control and manipulate new functionalities across their heterointerfaces \cite{Dagotto,Huijben,Zhuravlev,Gruverman,Garcia}. These materials, typically of the form ABO$_3$, exhibit strong correlations between the charge, spin, orbital and lattice degrees of freedom, leading to modulation of their device properties and are emerging as strong contenders for Beyond Moore technology. Interfaces coupling complex oxides exhibit novel functionalities, not present in their parent compounds \cite{Hwang}. These primarily stem from broken symmetries at the interface, which when carefully tailored, can lead to unique device functionalities. In this regard, SrRuO$_3$ is a canonical metallic system that has been studied extensively for its growth and physical properties and is widely used as an electrode in complex oxide devices \cite{Sanchez,Herranz,Bern,Aso,Koster}. Interestingly, SrRuO$_3$ undergoes a ferromagnetic transition at 140 K (T$_C$) and presents a distorted orthorhombic structure \textit{Pbnm} below 820 K. The microscopic origin of ferromagnetism is supported by the distortions of the oxygen octahedra that are coupled to the degree of hybridization between O $\textit 2p$ and the Ru $\textit 4d$ orbitals \cite{Vailionis1}.This strongly influences its magnetic and electronic properties on either side of the T$_C$
and provides an interesting opportunity to study the correlation between magnetism and transport in SrRuO$_3$ across its T$_C$ when integrated in an electronic device.    \\

In this work, we employ a nanoscale transport probe to study such complex correlation across an interface between SrRuO$_3$ on semiconducting Nb:SrTiO$_3$ (Nb:STO). Principally, this technique allows us to study any material system, whose electronic properties are susceptible to a temperature driven phase transition, while simultaneously allowing us to investigate its homogeneity across such buried interfaces with a semiconductor, at the nanoscale. This is an important approach for material systems such as complex oxides where small changes in the geometrical properties with thickness or temperature can produce large changes in the electric and magnetic properties. We perform a systematic experimental study to understand the effect of thickness and temperature on electronic transport across thin films of SrRuO$_3$ (t = 6, 8, 9 and 10 unit cells) with Nb:STO and observe an order of magnitude increase in the transmission in the same device, across its ferromagnetic phase transition. Surprisingly though, we find that this increase originates from an enhanced transmission at the interface between SrRuO$_3$ and Nb:STO, rather than in the bulk of SrRuO$_3$, and manifests as two different transport lengths at room temperature and below T$_C$. We combine these findings with density functional theory (DFT) calculations to quantitatively understand the role of interface reconstruction, octahedral distortions and exchange splitting in the observed features in electronic transport.\\

\section*{Results}

\subsection*{Film growth and device scheme}
We have grown high quality films of SrRuO$_3$ of varying thicknesses, 6, 8, 9 and 10 unit cells (u.c.) on 0.01 wt.\% Nb:STO substrates using Pulsed Laser Deposition (PLD). The substrates are chemically treated to ensure a uniform TiO$_2$ termination and characterized for their surface quality with atomic force microscopy (AFM) \cite{Koster1}. This process assumes critical importance because a clean, defect free starting surface is highly desirable for the growth of crystalline epitaxial films of SrRuO$_3$ and determines a uniform electronic transport across such a metal-semiconductor (M-S) interface. However, thermal annealing often causes local SrO terminations commonly occurring at the terrace edges of the annealed substrates. With unequal growth rate of SrRuO$_3$ on SrO and TiO$_2$ terminations, clear trenches along local SrO terminating sites are observed where SrRuO$_3$ grown is thinner than on TiO$_2$ terminations \cite{Koster}. Figure 1a,b shows the AFM image of the deposited films of different thicknesses (6 and 10 u.c.). The root-mean-square surface roughness of the 6 u.c. and 10 u.c. films, outside the trenches, are found to be 0.21 nm and 0.15 nm respectively. The grown SrRuO$_3$ films were characterized for their electronic transport and magnetic properties using standard van der Pauw method and superconducting quantum interference device (SQUID) magnetometry, respectively. The electrical resistivity studies on the films show a typical metallic behavior for all thicknesses. Resistivities of 6 and 10 u.c. SrRuO$_3$ films, measured from 300 K down to 15 K, are shown in Figure 1c. Their electrical properties are strongly affected by the film thickness, evidenced by a higher resistivity observed for the 6 u.c. film that decreases for the 10 u.c. film. A study of the temperature variation of resistivity shows a kink indicating the paramagnetic to ferromagnetic transition, with the corresponding T$_C$ (marked by black arrows) derived from the temperature dependence of the derivative of resistivity plots (not shown). The Curie temperature (T$_C$) determined for the 6 u.c. film is 135 K, and for 10 u.c. is 142 K. From SQUID magnetometry studies, the magnetic properties of the films were studied in an applied field of 0.1 T on field cooling for a temperature range between 5 K to 300 K (Figure 1d). The T$_C$ for the 6 u.c. and 10 u.c. films determined from these plots (derivative of magnetic moment with temperature, not shown) are found to be 135 K and 142 K respectively, and demonstrates the coupled magnetic and electric properties in SrRuO$_3$. The high value of the saturation magnetic moment obtained from experiments arises due to the strained films on (001) SrTiO$_3$, which decreases the orbital overlap via increased bond angle in Ru-O bonds, in such ultrathin films of SrRuO$_3$. This has been reported earlier in SrRuO$_3$ films \cite{Grutter, Grutter1}. Thus, through a combination of surface characterization technique and physical property measurements, we establish the high quality of the films used in our work.\\

In order to probe electronic transport of the deposited SrRuO$_3$ films, they are patterned with standard UV lithography and further ion beam etched to be defined as devices as shown in Figure 1e. Our experimental nanoscale transport probe is known as ballistic electron emission microscopy (BEEM) (Figure 1c) which uses the scanning tunneling microscope (STM) tip to inject a distribution of hot electrons with energies higher than the Fermi energy (E$_F$) \cite{Kaiser,Bell}. In this study, we use a modified Ultra High Vacuum (UHV) STM system from RHK Technology and the measurements are performed at 300 K and 120 K using PtIr metal tips. An electrical contact between the STM tip and SrRuO$_3$ measures the tunnel current. The device design necessitates an additional contact at the rear-end of Nb:STO, for the collection of the transmitted electrons. These injected hot electrons, driven by the applied bias \textit{V$_T$} with respect to the Fermi level of the metal, travel across the metallic base to reach the M-S interface; if their energy is sufficient to overcome the Schottky barrier, they can be collected into the semiconductor conduction bands (Figure 1f). To obtain the BEEM spectra, the STM tip is held at a fixed location while \textit{V$_T$} is varied within a voltage range and the collector current is recorded in the constant current mode. As these hot electrons propagate through the SrRuO$_3$  films, they undergo scattering, which influences the collected current. The collected electrons in Nb:STO are primarily those whose momentum parallel to the interface is conserved, the rest are back-scattered into the SrRuO$_3$ base layer. These electrons serve as an important parameter to study strong correlations in such metallic oxide thin films and provide information on the homogeneity of the interface electronic structure with high spatial resolution \cite{Prietsch,Kanel,Parui1,Haq}. The determination of the relative strengths of such scattering processes (elastic/inelastic) enables a comprehensive understanding of electron transport in complex oxides and allows for designing of new oxide based electronic devices. \\

\subsection*{Thickness and temperature dependent BEEM transmission}
The hot electron distribution encounters elastic and inelastic scattering while transmitting through the SrRuO$_3$ metallic base. In this process, the momentum distribution at the M-S interface is broadened, as compared to the injection interface, adding to the reduction of the collected BEEM current (I$_B$) \cite{Subir}.These effects get more pronounced as the thickness (d$_{SrRuO_3}$) of the base layer is increased. The normalized BEEM transmission follows an exponential behavior as described by \cite{Kaiser}:

\begin{equation}
\frac{I_B}{I_T} = C \times exp\left[-\frac{d_{SrRuO_3}}{\lambda_{SrRuO_3}(E)}\right]
\end{equation}

where $\lambda_{SrRuO_3}(E)$ is the energy-dependent hot electron attenuation length, I$_B$ is the collected BEEM current, I$_T$ is the injected tunnel current and C is a constant that is the kinematic transmission factor representative of electron transport across the M-S interface. Figure 2 shows the BEEM transmission data for various thickness of SrRuO$_3$ thin films, in their paramagnetic (300 K) and ferromagnetic (120 K) phases. First we discuss the observation for the paramagnetic phase (Figure 2a). The  transmission for all film thicknesses is low up to a certain bias voltage beyond which, the transmission increases rapidly. This onset in bias voltage corresponds to the local Schottky barrier height (SBH) at the M-S interface. In accordance to the Bell-Kaiser (B-K) model, a plot of the square root of BEEM current versus the bias voltage yields the value of SBH, which in this case is 1.13$\pm$0.03 eV (Inset, Figure 2a) \cite{Rana}. It matches well with our previously extracted values for such an interface \cite{Roy}. As the thickness of SrRuO$_3$ films is increased, a decrease in transmission is observed. However, the local SBH extracted from measurements on films of different thicknesses are found to be the same, thus providing yet another proof of a high quality growth of atomically sharp M-S interfaces. The representative BEEM spectra shown for each thickness are averages of several spectra taken at different devices and locations. \\

Electronic transport across the ferromagnetic phase transition in SrRuO$_3$ thin films was studied across its interface with Nb:STO, in the same set of devices, by performing similar BEEM measurements at 120 K. The characteristic BEEM spectra are shown in Figure 2b. Similar to our earlier observations, the BEEM transmission progressively decreases with increasing thickness of the films. We find that the hot electron transmission increases by an order of magnitude as compared to its corresponding values at 300 K. The extracted SBH at 120 K, at these interfaces, is found to be 1.14$\pm$0.03~eV (Inset, Figure 2b), which is the same as that obtained in its paramagnetic phase. One can thus safely discard the origin of the enhanced BEEM transmission in SrRuO$_3$ at 120 K, due to differences in the SBH across its magnetic phase transition.\\

\subsection*{Influence of interface and film thickness on electronic transport} 
An exponential fit of the BEEM transmission for varying thicknesses of SrRuO$_3$ films allows us to extract the (energy dependent) transport length scale known as attenuation length (Equation 1). Above E$_F$, the transmission of hot electrons in SrRuO$_3$ films is mostly governed by inelastic scattering, originating from the availability of increased phase space for hot electrons to decay into \cite{Banerjee,Prietsch}. The extracted attenuation length of the carriers reflects the combined effects of inelastic, elastic as well as quasi-elastic scatterings. We plot the transmissions for the paramagnetic and ferromagnetic phase of SrRuO$_3$ at a certain bias voltage of -2 V, shown in Figure 3a. An important observation is that the BEEM transmission starts to deviate appreciably at 120 K from that at 300 K, with decreasing thickness. This has not been observed earlier in other metallic systems using BEEM. Extrapolation of the data for zero SrRuO$_3$ thickness shows two orders higher transmission, at the interface for ferromagnetic SrRuO$_3$. An exponential fit using Equation 1, yields an attenuation length of 1.6$\pm$0.2 u.c. in the paramagnetic phase and 0.88$\pm$0.4 u.c. in the ferromagnetic phase at -2 V. In Figure 3b, we plot the energy dependence of the attenuation length by fitting the data obtained in Figure 2 with Equation 1 for energies above the SBH. We find that the attenuation length decreases with increasing energy, consistent with the enhancement of the density of states (DOS) at higher energies \cite{Guedes}. What is surprising is that in-spite of the enhanced transmission in the ferromagnetic phase, the attenuation length in SrRuO$_3$ is shorter, for all energies at 120 K, than at 300 K. This apparent conundrum is explained by our \textit{ab initio} calculations, taking into account the characteristics of geometries and electronic structures at the interface between SrRuO$_3$ and SrTiO$_3$ along with the quasiparticle renormalization of attenuation length.\\

\subsection*{\textit{Ab initio} study of SrRuO$_3$/SrTiO$_3$(001) interfaces} 
We have performed \textit{ab initio} density functional calculations with a specific focus on correlating the structural and electronic properties at the interface to the BEEM transmission and to understand the origin of the flipping of the attenuation lengths at 120 K and 300 K. In order to simulate SrRuO$_3$ thin films grown on SrTiO$_3$ (001) substrates, we fix the lattice parameter $a$ to the experimental value of the SrTiO$_3$ substrate. 3, 6 and 9 unit cells of SrRuO$_3$  have been considered in the low temperature ferromagnetic and room temperature (RT) paramagnetic phases. As a paramagnetic phase with disordered local moments is difficult to simulate within our computational technique, we study the nonmagnetic phase as the RT phase and ferromagnetic phase calculated at 0 K will represent the low temperature (LT) phase. We show in Figure \ref{fig:figure1TH}a the supercell with 6 layers of SrRuO$_3$ on 3 layers of SrTiO$_3$.
In the following discussions, we will follow the notation shown in Figure \ref{fig:figure1TH}a, where we denote STOL (or SROL) with L=1,2,..., the L-th layer from the SrTiO$_3$/SrRuO$_3$ interface. \\

To determine the energy dependent attenuation length $\lambda^{LDA}(E)$,
we calculate the mean value over the bands and k-points, at a given energy $E$,
of the product of the modulus of the group velocity along the $z$ axis $v^z_{n\textbf{k}}$ and the lifetime $\tau_{n\textbf{k}}$, which is expressed as:
\begin{equation}\label{ATT_LDA}
\lambda^{LDA}(E)=\langle v^z_{n\textbf{k}}\tau_{n\textbf{k}}\rangle \propto \frac{\langle\frac{1}{\hbar}\frac{\partial\varepsilon_n(\textbf{k})}{\partial{k_z}}\rangle}{\int_0^E N(\varepsilon)d\varepsilon}
\end{equation}
where the lifetime is proportional to the inverse of the cumulative density of states from the Fermi level to the energy $E$\cite{Zhukov}. We have only considered the z-component, as the BEEM transmissions across the M-S interface are measured along this direction. As seen in Figure \ref{fig:figure1TH}b, the attenuation length $\lambda^{LDA}(E)$ is almost constant in the range between 1.6 and 2.2 eV (above E$_F$) and is greater at 120 K than at RT. In recent works \cite{Renormalization,Renormalization2}, the importance of dynamical correlations in SrRuO$_3$ have been discussed with the aid of dynamical mean field theory. Whereas the dynamical correlations and hence, the temperature dependent quasiparticle weight Z are beyond the scope of LDA, we note that this weight is 1 in LDA and is a multiplicative factor of the spectrum and the velocity. Assuming the same quasiparticle weight for the $e_g$ manifold and neglecting its influence on $\tau$, we get a reduction of the attenuation length, which is more pronounced at lower temperatures. In Figure \ref{fig:figure1TH}b  we plot the attenuation length multiplying $\lambda^{LDA}$ by $Z$ and observe that under these assumptions, the attenuation length at RT is greater than at 120 K, in agreement with the experimental findings. \\

\subsection*{Variation of the electronic and structural properties of the interface at LT and RT} 
To understand the experimentally observed enhancement in the interface transmissions at 120 K and 300 K, we calculate the electronic and geometric structures of SrRuO$_3$/SrTiO$_3$ interfaces. Considering the variation in the geometric and magnetic properties, we can divide the SrRuO$_3$ thin films in three regions along the c-axis: surface, inner layer and interface region. Our calculated densities of states of SrRuO$_3$ along with that of SrTiO$_3$ are shown in Figure \ref{fig:figure4TH}a for 9 u.c. of SrRuO$_3$ in the LT ferromagnetic phase and the RT nonmagnetic phases. It is observed that between 1~eV and 2.2~eV (above E$_F$), the DOS in SrRuO$_3$ is dominated by $e_g$ electrons whereas in SrTiO$_3$, the $t_{2g}$ electrons dominate with a larger DOS between 1.2 and 1.4~eV (above E$_F$) in the nonmagnetic phase of SrRuO$_3$. The exchange split $e_g$ state produces a smaller DOS at LT between 1.2 and 1.4~eV (above E$_F$). While we observe the differences between LT and RT phases in all SRO layers, SRO1 layer at the interface needs an extra attention. A distinct feature of $d_{x^{2}-y^{2}}$ character (marked by the shaded area) is observed for RT phase in the SRO1 layer along with the renormalization of $e_g$ DOS between 1.2 and 1.3 eV (above E$_F$). This is not an energy shift but is due to an enlargement of the bandwidth. Our calculated hopping parameters, for example, for the 6 u.c. at RT are 580~meV between the $x^2$-$y^2$ orbitals (corresponding Wannier function shown in Figure \ref{fig:figure1TH}a) in the SRO1 layer and 540-543~meV for other layers. Also, the hopping mismatch between interface (575 meV) and inner layers (554-561 meV) is reduced at LT. Thus, we can clearly see that the mismatch between the interface and inner layers is larger at RT. We further note from our DOS calculations that the mismatch of the electronic states at the interface between SrRuO$_3$ and SrTiO$_3$ is larger at RT than at LT. Such elastic scattering effects will lead to a reduction in the BEEM transmission at RT. All these differences in the electronic structures of the two phases strongly suggest the importance of the prefactor C in Equation 1, which influences the BEEM transmission at LT and RT phases.\\

The difference in the characteristics at the interface for LT and RT phases is quite evident also in the structural properties shown in Figure \ref{fig:figure4TH}b. The interface reconstruction makes this aspect more evident in the in-plane bond angles compared to the out-of-plane bond angles. For a slab of reasonable thickness (9 u.c. in our case), one can quantitatively identify the surface, inner layers and interface regions by the evolution of the Ru-O-Ru bond angle in the \textit{ab} plane along the c-axis as shown in Figure \ref{fig:figure4TH}b. For the 9 u.c. case, a large region of inner layers (properties similar to bulk) exists while this region is not distinctly observed in the 3 u.c. case. Due to the interface reconstruction, the interface layer SRO1 is characterized by a Ru-O-Ru in-plane bond angle having a value between the SrRuO$_3$ and SrTiO$_3$ bulk values. The properties of the inner layers are closer to the bulk. At the surface, we always obtain a bond angle larger than the bulk. Apart from the general differences between bond angles for different regions of the slab, an important observation can be made regarding the bond angles for the LT and RT phases. \\

It is clearly seen that the difference in bond angle between SRO1 and other SRO layers is larger for the RT phase compared to the LT one. For example, in the 6 u.c. case, the difference in the bond angle between SRO1 and SRO5 is 1.5$^{\circ}$ for the LT phase, while it is 3.0$^{\circ}$ for the RT phase (Figure \ref{fig:figure4TH}b). This influences the hopping of electrons (as evident from the calculated hopping parameters mentioned above) and hence the transport properties at the interfaces \cite{Autieri14}. Therefore, we can conclude that the differences in the geometries and electronic structures between SRO1 and other SRO layers are less dramatic in the LT phase compared to the RT one, which may effectively lead to an enhanced transmission of hot electrons across the interface at LT. The larger mismatch at RT between interface and inner layers will yield a smaller transmission factor (C factor in Equation 1) in the RT phase. Such a substantial influence of geometric and structural changes at the interface leading to strong temperature dependent transport characteristics across the magnetic phase transition in SrRuO$_3$ is a remarkable finding, not observed before in any oxide heterostructure.\\

\section*{Discussion}
In conclusion, we have studied the BEEM transmission through thin films of SrRuO$_3$ of various thicknesses, interfaced with SrTiO$_3$ across its magnetic transition. The increase in BEEM transmission at 120 K compared to RT is accompanied by a surprising flipping of the attenuation length. Our first principles calculations indicate strong geometrical reconstructions at the interfaces characterized by Ru-O-Ru bond angles along with distinct features in electronic structures of the Ru-d orbitals, both of which are strongly dependent on the magnetic phases. As a consequence, the importance of the transmission factor in the expression for BEEM current is realized to explain the experimentally observed BEEM transmission data. Moreover, the inclusion of the temperature dependent quasiparticle weight for the calculation of the attenuation length correctly describes the experimental observation at 120 K and RT.

The evolution of such unique features in electronic transport in complex oxide heterostructures, across a magnetic phase transition, opens new possibilities to manipulate the electronic and spin degrees of freedom by a selective choice and tailoring of their interfaces. This approach of tuning heterointerfaces by coupling structural, electronic and magnetic properties can be extended to other material systems with promising prospects for future oxide electronic and spintronic devices.\\

\section*{Methods}

\textit{Experimental details} We grow SrRuO$_3$ films of different thickness by pulsed laser deposition. All the devices were grown on [001] Nb doped SrTiO$_3$ substrates with 0.01 wt.$\%$ of Nb. The Nb:STO substrates used for deposition were treated chemically to obtain uniform TiO$_2$ terminations, and were annealed at 960$^{\circ}$ C which rendered a clean surface with uniform straight terraces. The cleanliness and uniformity of the interface becomes critical in electronic transport across the heterointerface. The films were grown using PLD with a KrF excimer laser with a repetition rate of 1 Hz at an oxygen pressure of 0.13 mbar at 600$^{\circ}$ C. Their corresponding thicknesses were determined \textit{in-situ} by reflection high energy electron diffraction. Electrical resistivities of the films were obtained by standard van der Pauw method within a temperature range of 15 K to 300 K. Magnetic measurements were performed using a Quantum Design MPMS XL-7 SQUID. In order to fabricate devices, patterns of 100x150 $\mu$ m$^2$ areas were defined using standard ultraviolet (UV) lithography technique where 100 nm of Au was deposited for making the top contact, providing the pathway for the STM current. Thereafter, device areas of 200x300 $\mu$ m$^2$ were defined with the Au pads on top of it, and the samples were Ion Beam Etched (IBE) with a pre-calibrated etch-stop. This removed the SrRuO$_3$ from the unpatterned regions isolating the adjacent devices on the sample. An Ohmic back contact was created by depositing Ti (70nm)/Au (70nm) on the rear side of the Nb:STO substrate for collecting the BEEM current. The photoresist was lifted-off carefully and the topography of SrRuO$_3$ was checked using AFM. \\

\textit{Computational details} We perform spin polarized first-principles density functional calculations within the LSDA (Local Spin Density Approximation) \cite{Kohn64} by using the plane wave VASP \cite{VASP} DFT package and the Perdew-Zunger \cite{Perdew} parametrization of the Ceperly-Alder data \cite{Ceperley} for the exchange-correlation functional. The choice of LSDA exchange functional is suggested in a recent paper \cite{Etz12} where the Generalized Gradient Approximation was shown to perform worse than LSDA for SrRuO$_3$. The interaction between the core and the valence electrons was treated with the projector augmented wave (PAW) method \cite{BlochlPAW} and a cutoff of 430~eV was used for the plane wave basis.The computational unit cells are constructed as supercells with three SrTiO$_3$ octahedra having a 12 {\AA} of vacuum. Convergence was checked with respect to the thickness of the vacuum layer. Depending on the thickness of SrRuO$_3$ films, 3, 6 and 9 octahedra of SrRuO$_3$ are placed on top of  SrTiO$_3$. For Brillouin zone integrations, a 8$\times$8$\times$1 k-point grid is used for geometry relaxation and a 10$\times$10$\times$2 k-point grid for the determination of the density of states (DOS). In all cases, the tetrahedron method with Bl\"{o}chl corrections \cite{BlochlCORR} is used for the Brillouin zone integrations. We optimize
the internal degrees of freedom by minimizing the total energy to be less than 10$^{-5}$~eV and the Hellmann-Feynman forces to be less than 7~meV/{\AA}. The Hubbard $U$ effects at the Ru and Ti sites were included in the LSDA+$U$ \cite{Anisimov91,Anisimov93} approach using the rotational invariant scheme proposed by Liechtenstein \cite{Anisimov95}.
We have used $U$=5~eV and $J_H$=0.64~eV for the Ti 3$d$ states \cite{Pavarini04} while  $U$=0.30~eV and $J_H$=0.05~eV are considered for the Ru 4$d$ states. In the literature, a value of the Coulomb repulsion of 0.50~eV for ruthenates \cite{Etz12} was suggested to reproduce the magnetization of the bulk, but we use a slightly smaller value because a smaller magnetization is found for the thin films\cite{Tian07}. To extract the character of the electronic bands at low energies, we used the Slater-Koster interpolation scheme based on Wannier functions.\cite{Marzari97,Souza01} Such an approach is
applied to determine the real space Hamiltonian matrix elements in the e$_g$-like Wannier function basis using a 8$\times$8$\times$2 k-point grid. After obtaining the Bloch bands in density functional theory, the Wannier functions are constructed using the WANNIER90 code.\cite{Mostofi08} Starting from an initial projection of atomic $d$  basis functions belonging to the e$_g$ manifold and centered on Ru sites, we get the two e$_g$-like Wannier functions. The group velocity is obtained using a 49$\times$49$\times$49 k-point grid. We got qualitatively similar results for the Fig. \ref{fig:figure1TH}b using the quasiparticle weight Z calculated in different references\cite{Renormalization,Renormalization2}. \\

\noindent

\section*{Acknowledgements}

We thank B. Noheda and T. T. M. Palstra for use of the Pulsed Laser Deposition system. Technical support from J. Baas and J. G. Holstein is thankfully acknowledged. This work is supported by the Netherlands Organization for Scientific Research NWO-FOM (nano) and the Rosalind Franklin Fellowship program. C.~A. and B.~S. acknowledge financial support from Carl Tryggers Stiftelse (grant no. CTS 12:419 and 13:413) and supercomputing allocation by Swedish National Infrastructure for Computing.  \\

\section*{Author contributions statement}

S. R. carried out the device fabrication, all experimental measurements and contributed to the analysis of the data. T. B. assisted in experimental planning, contributed to the data analysis and supervised the overall work. C. A. did the theoretical calculations and took part in the analysis of theoretical data. B. S. supervised the theoretical part of the project and took part in the analysis of theoretical data. All coauthors extensively discussed the results and wrote the paper. \\

\section*{Additional information}

Competing financial interests: The authors declare no competing financial interests.\\

\renewcommand{\figurename}{{\bf Figure}}
\begin{figure}[h]
\includegraphics[scale=0.70]{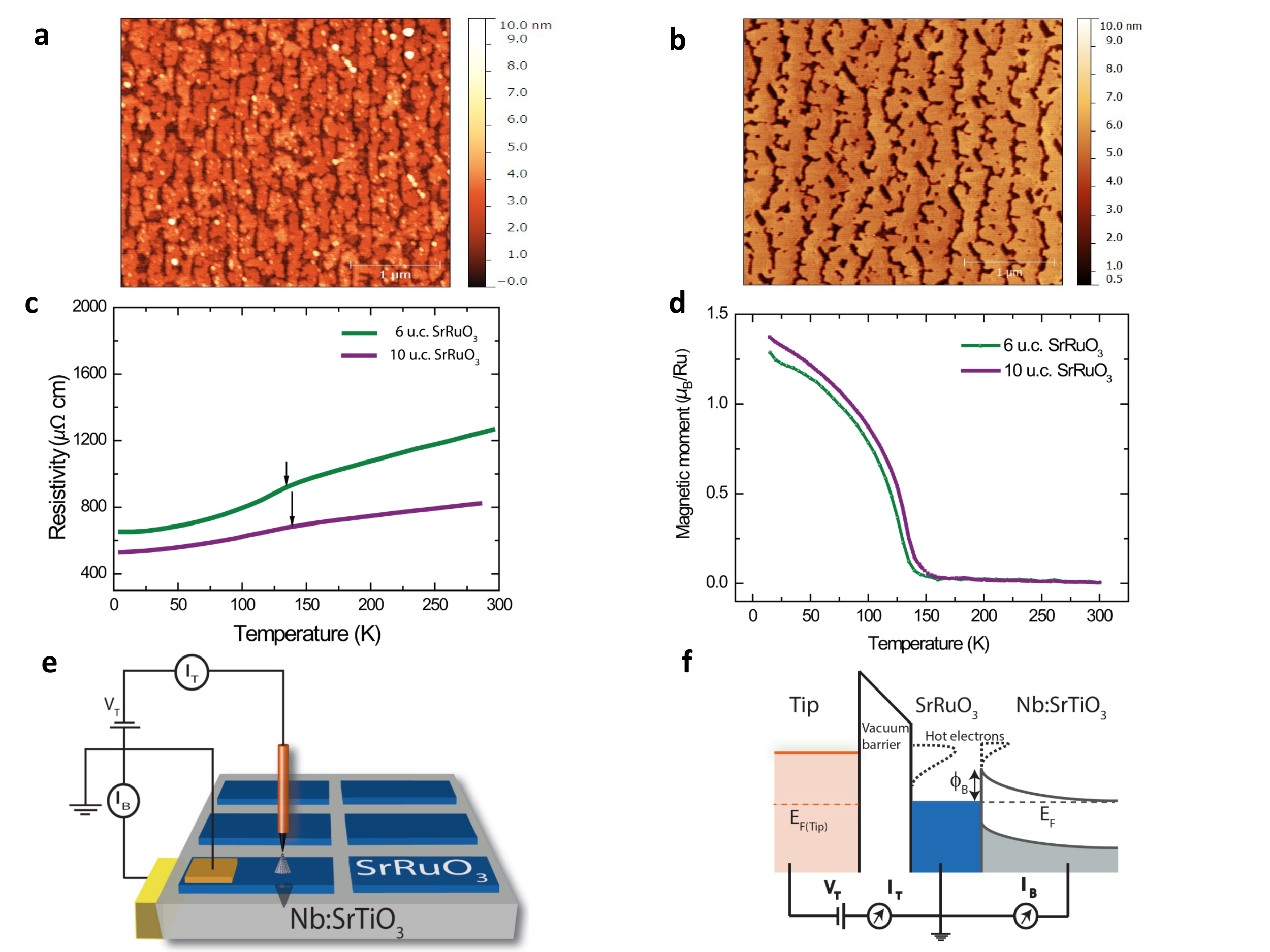}
\caption{\label{1.} {\bf Atomic force microscopy (AFM) images of grown films and device design.} {\bf(a)} 6 u.c. of SrRuO$_3$ grown on Nb:STO. Clearly, the presence of local SrO terminations of the substrate is observed as dark patches where SrRuO$_3$ growth rate is lower than on TiO$_2$ terminations of the substrate. {\bf(b)} Topography image after 10 u.c. of SrRuO$_3$ growth. As the film grows thicker, it starts covering the local trenches with lower SrRuO$_3$ growth ultimately replicating the substrate morphology. We observe a dependence of thickness on the uniform coverage of SrRuO$_3$ on the substrate. {\bf(c)} Temperature dependence of resistivity of 6 and 10 u.c. SrRuO$_3$ films. The T$_C$ obtained from the kink in the curves indicates that it decreases with decreasing film thickness. Black arrows indicate the T$_C$ of the films, which is 135 K and 142 K respectively for the 6 u.c and 10 u.c of SrRuO$_3$. Also, 10 u.c. SrRuO$_3$ is more metallic than 6 u.c. SrRuO$_3$. {\bf(d)} Magnetic moment dependence on temperature is recorded on field cooling in a magnetic field of 0.1 T. Extracted T$_C$ for 6 u.c. and 10 u.c. SrRuO$_3$ films are 135 K and 142 K, respectively. These values are consistent with the findings from temperature dependent resistivity measurements.  {\bf(e)} Schematic design of SrRuO$_3$/Nb:STO device and measurement scheme. Electrons are injected by the STM tip and BEEM current is collected from Nb:STO. {\bf(f)} Energy band profile of the measurement scheme. The Schottky barrier at the SrRuO$_3$/Nb:STO interface acts as an energy filter for the transmitted hot electrons.}
\end{figure}

\renewcommand{\figurename}{{\bf Figure}}
\begin{figure}
\includegraphics[scale=0.10]{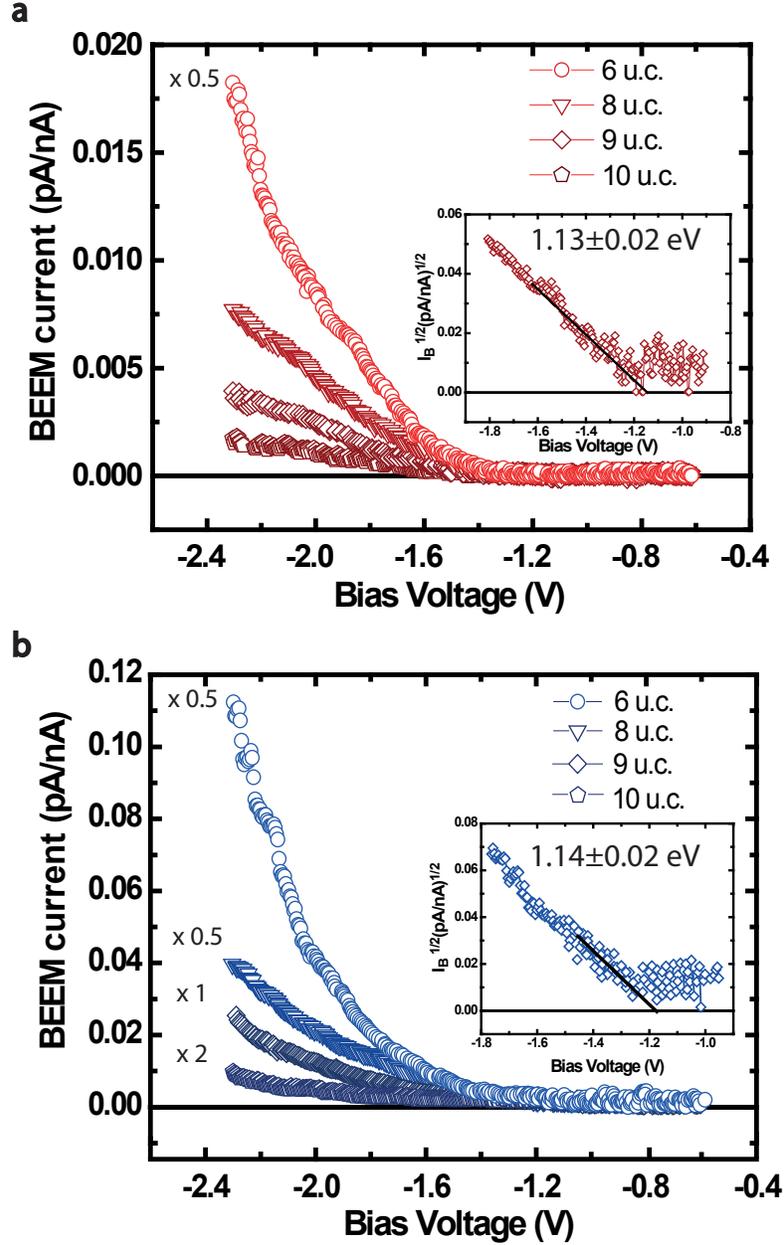}
\caption{\label{2.} {\bf Thickness dependent BEEM transmission in SrRuO$_3$/Nb:STO devices at 300 K and in the ferromagnetic phase at 120 K}. {\bf (a)} A plot of BEEM current vs. bias voltage is shown for different thicknesses of SrRuO$_3$ at 300 K. Clearly, the transmission decreases progressively with increasing film thickness. The curve for 6 u.c. SrRuO$_3$ has been multiplied by 0.5 factor. The inset shows representative local Schottky barrier height (SBH) extracted by the Bell-Kaiser model, for the interface of metallic SrRuO$_3$ and semiconducting Nb:SrTiO$_3$. For all the thicknesses, it is found to be 1.13$\pm$ 0.02 eV. The error bars reflect the standard deviation of the measured onset of the BEEM current at a location. {\bf (b)} The plot of BEEM current vs. bias voltage shows an thickness dependence, similar to what we observe at 300 K. An increased BEEM transmission for all thicknesses is observed compared to 300 K. The curves for 6 u.c. and 8 u.c. of SrRuO$_3$ have been multiplied by 0.5 and the 9 u.c. SrRuO$_3$ curve by a factor 2. The extracted local SBHs for the devices are 1.14$\pm$ 0.03 eV, similar to what was observed at 300 K.}
\end{figure}

\renewcommand{\figurename}{{\bf Figure}}
\begin{figure}
\includegraphics[scale=0.06]{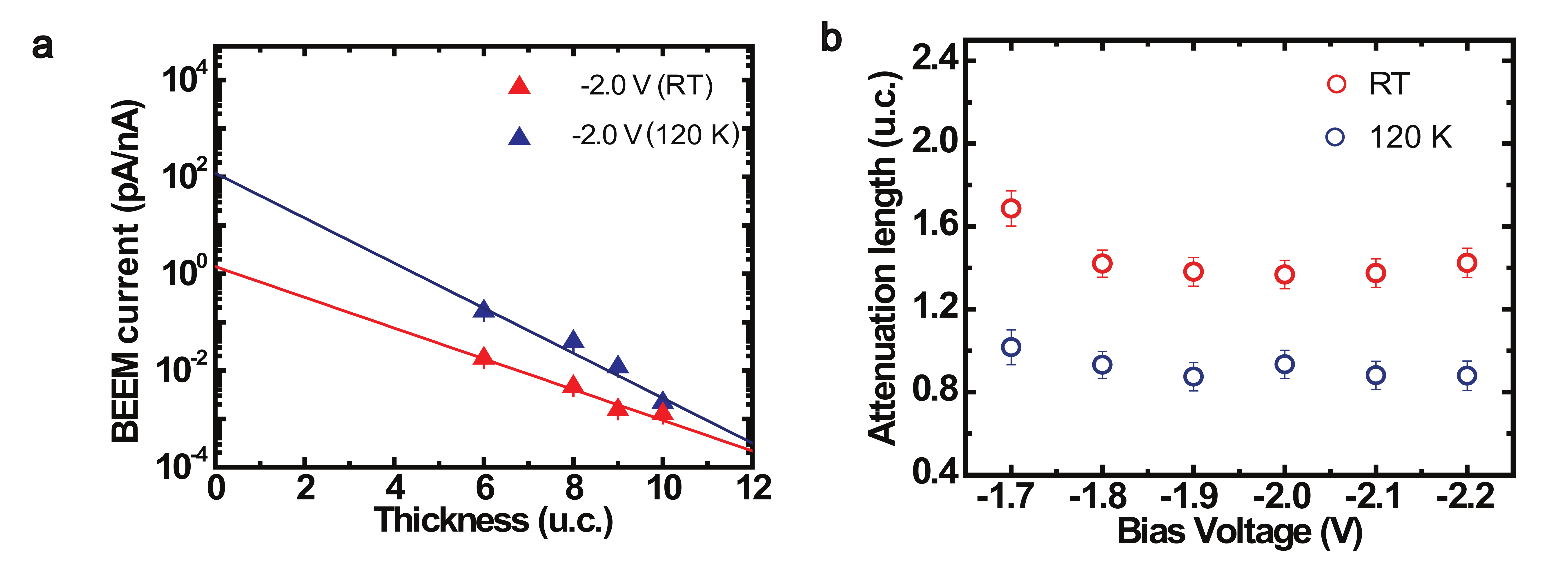}
\caption{\label{3.} {\bf Plot of attenuation length with applied bias and energy dependence}. {\bf (a)} The BEEM transmission at -2 V is plotted against film thicknesses for both at 300 K (RT) and at 120 K. An exponential fit (solid lines) show the dependence of BEEM current with film thickness. An extrapolation of the transmission indicates that the interface transmission has increased by two orders of magnitude at 120 K. {\bf (b)} Energy dependence of attenuation length at 120 K and 300 K is plotted. Contrary to the higher BEEM transmission across SrRuO$_3$/Nb:STO devices at 120 K, the corresponding attenuation length is consistently shorter than the attenuation length measured at 300 K. The error bar in attenuation length comes as a fitting error from the deviation of the best fit.}
\end{figure}

\renewcommand{\figurename}{{\bf Figure}}
\begin{figure}[!ht]
\centering
\includegraphics[scale=0.4]{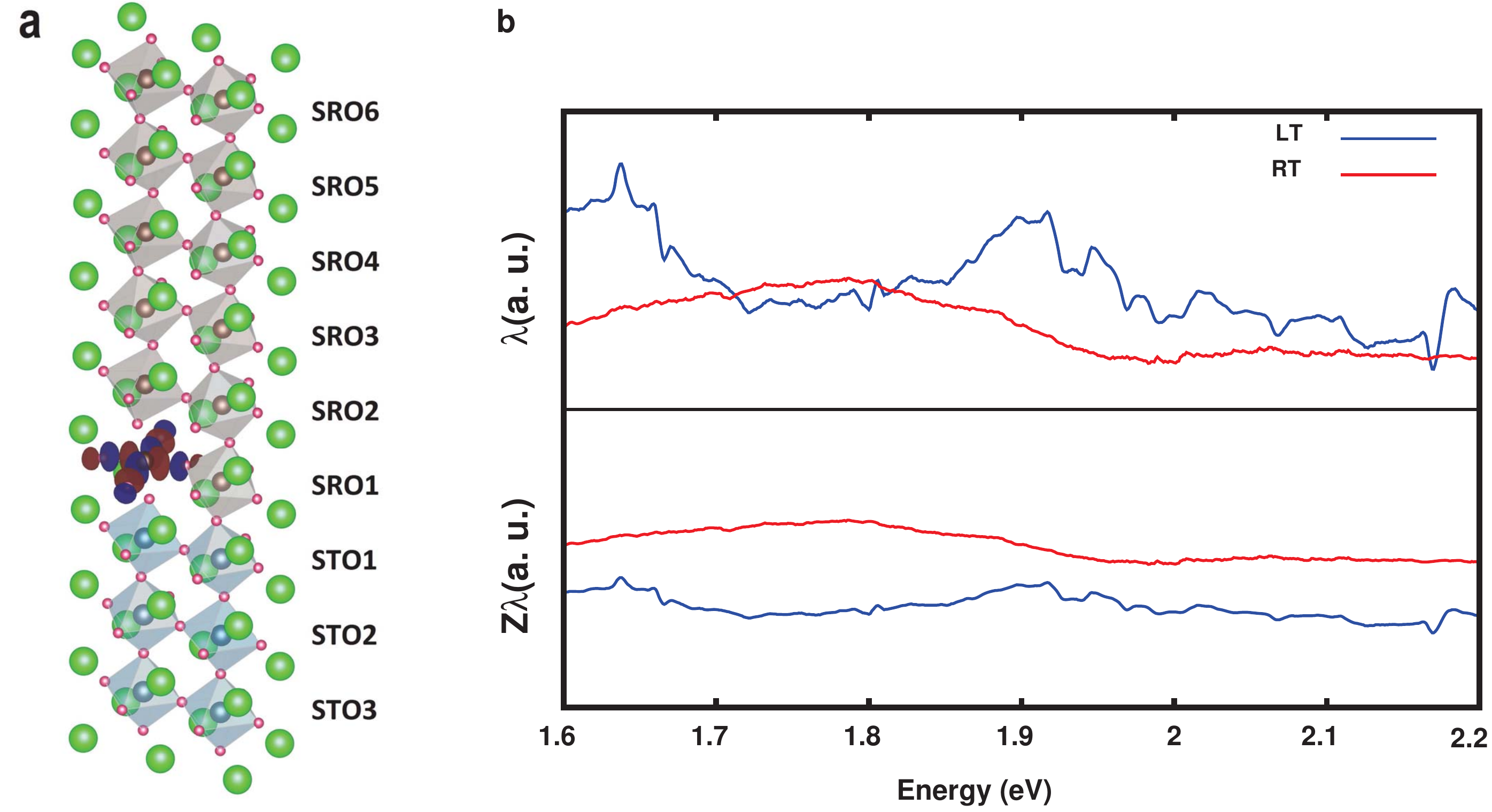}
\caption{{{\bf Side view of the slab and calculated attenuation length.} {\bf (a)} Side view of the slab with 3 layers of SrTiO$_3$ and 6 layers of SrRuO$_3$.
Sr, O, Ru and Ti atoms are shown as green, fuchsia, grey and blue balls respectively. In the SRO1 layer, we show the Wannier function corresponding to the $d_{x^{2}-y^{2}}$ orbital where red and blue contours are for isosurfaces of identical absolute values but opposite signs. {\bf (b)} Attenuation length calculated within LDA without and with the quasiparticle weight Z. The top and bottom panels show $\lambda^{LDA}$ and Z$\lambda^{LDA}$ respectively in the energy range between 1.6 and 2.2 eV (above E$_F$) for the case with 6 layers of SrRuO$_3$. This energy range is the relevant experimental energy range where the attenuation length was measured. The LT (RT) phase is shown in blue (red).}}
\label{fig:figure1TH}
\end{figure}

\renewcommand{\figurename}{{\bf Figure}}
\begin{figure}[!ht]
\centering
\includegraphics[scale=0.55]{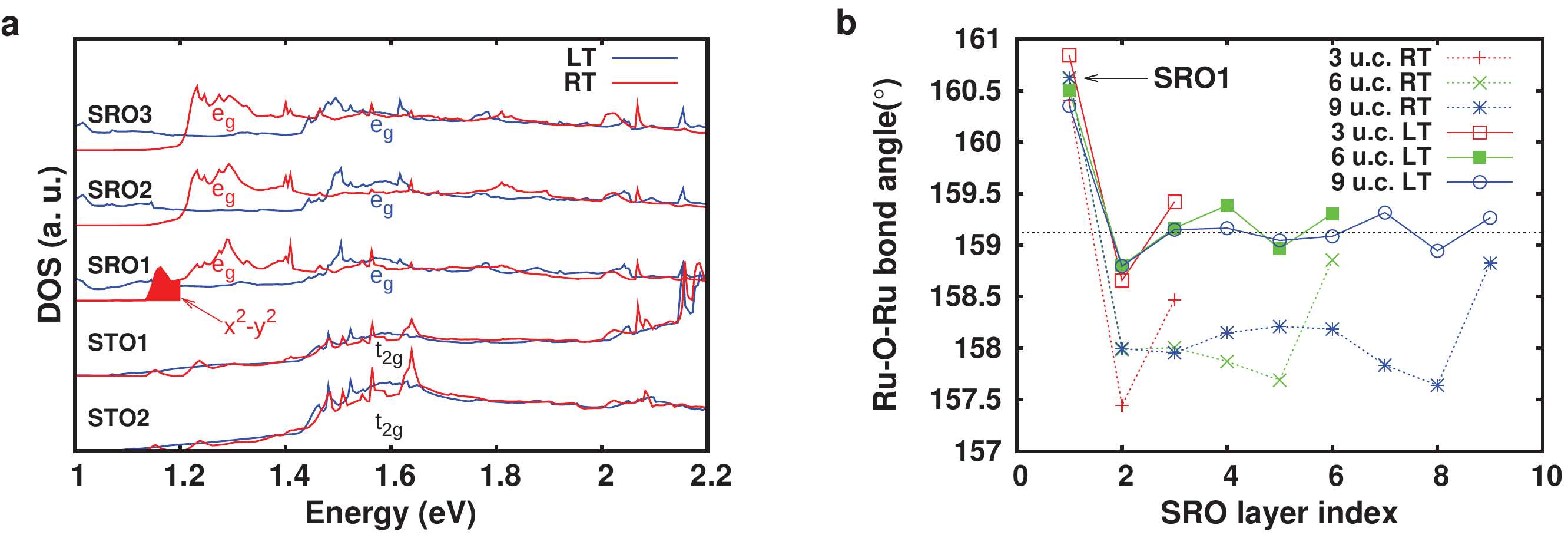}
\caption{ {\bf Layer-resolved electronic and structural properties.} {\bf (a)} Layer projected DOS (spin-up + spin-down) closer to the interface from STO2 to SRO3 for the 9 u.c. case. The LT (RT) phase is shown in blue (red). The shaded DOS below 1.2~eV in the RT SRO1 layer has $x^2$-$y^2$ character. The energy range is from 1.0~eV to 2.2~eV (above E$_F$). {\bf (b)} Ru-O-Ru bond angle in the \textit{ab} plane as a function of the layer index for LT (solid lines) and RT (dashed lines) phases. The red, green and blue lines represent respectively the 3 u.c., 6 u.c. and 9 u.c. cases. The black dashed line represents the calculated LT bulk value in the same volume setup. The values for the SRO1 layer are indicated by an arrow.
}
\label{fig:figure4TH}
\end{figure}

\end{document}